\documentstyle[seceq]{ptptex}
\newtheorem{theorem}{Theorem}
\newtheorem{lemma}[theorem]{Lemma}
\newtheorem{coro}[theorem]{Corollary}

\newtheorem{formula}[theorem]{Formula}
\newtheorem{formulae}[theorem]{Formulae}

\title{%        %You can use \\ for explicit line-break
Criticality and Averaging in Cosmology
}
\author{Masayuki {\sc Tanimoto}\footnote{JSPS Research Fellow. 
}}

\inst{%         %Affiliation, neglected when [addenda] or [errata]
Yukawa Institute for Theoretical Physics, 
Kyoto University, Kyoto 606-8502, Japan}

\recdate{ August 13, 1999 }

\abst{ We propose comparing cosmological solutions in terms of their
  total spatial volumes $V(\tau)$ as functions of proper time $\tau$,
  assuming synchronous gauge, and with this intention evaluate the
  variations of $V(\tau)$ about the
  Friedmann-Lema\^{\i}tre-Robertson-Walker (FLRW) solutions for dust.
  This can be done successfully in a simple manner without solving
  perturbation equations. In particular, we find that first variations
  vanish with respect to all directions which do not possess homogeneity
  and isotropy preserving components; in other words, every FLRW
  solution is a {\it critical point} for $V(\tau)$ in the properly
  restricted subspace of the space of solutions. This property may
  support a validity of the interpretation of the FLRW solutions as
  constituting an averaged model. We also briefly investigate the second
  variations of $V(\tau)$.  }

\begin{document}

\def\pubnum{99--46}

\maketitle

\def\A{{\cal A}}
\def\B{{\cal B}}
\def\R{{\bf R}}
\def\d{{\rm d}}
\def\goes{\rightarrow}
\def\rcp#1{{1\over #1}}
\def\reff#1{(\ref{#1})}
\def\bra#1{\left[ #1 \right]}
\def\brace#1{\left\{ #1 \right\}}
\def\paren#1{\left( #1 \right)}
\def\G{{\cal G}}
\def\x{{\bf x}}
\def\tc{\tau_{0{\rm c}}}
%Metrics
\def\hh{{h}}
\def\gg{{\rm g}}
\def\uh#1#2{\hh^{#1#2}}
\def\dh#1#2{\hh_{#1#2}}
\def\ug#1#2{\gg^{#1#2}}
\def\dg#1#2{\gg_{#1#2}}
\def\uug#1#2{\tilde{\gg}^{#1#2}}
\def\udg#1#2{\tilde{\gg}_{#1#2}}
\def\udh#1#2{\tilde{\hh}_{#1#2}}
\def\sR{{}^{(3)}\!R}
\def\uK#1#2{K^{#1#2}}
\def\dK#1#2{K_{#1#2}}

\def\daa{\frac{\dot a}{a}}
\def\daas{\left( \frac{\dot a}{a} \right)^2}
\def\ka{\frac{k}{a^2}}
\def\gamsec{\gamma^{(2)}{}}
\def\lamsec{\lambda^{(2)}{}}

\section{Introduction}

The most important practical models in relativistic cosmology are the
so-called Friedmann-Lema\^{\i}tre-Robertson-Walker (FLRW) models, which
assume the homogeneity and isotropy of spatial geometry and matter
distribution.  Since our universe seems homogeneous and isotropic over a
large scale (typically the size of a supercluster), it may be natural to
expect that such a model provides a good approximation for the real
universe, describing averaged properties of the geometry and matter
distribution in it. However, \cite{El} justification of such an
expectation should be considered carefully, since Einstein's equation is
highly nonlinear, and thus no simple averaging procedure applied to
spatial geometry and matter commutes with time evolution.

The averaging problem in general concerns how one can modify Einstein's
equation for averaged variables to be compatible with the time evolution
of the original inhomogeneous data for the variables. To analyse the
problem one usually has to determine an averaging procedure. In fact,
many averaging procedures have been proposed,\cite{Fu}\tocite{Bu}
%\cite{Fu,Zal,CM,BE,RSKB,MAB,IHK,Boer,SHT,Bu} 
but the dynamical properties of the resulting averaged variables are
often unclear, and little firm agreement is seen among them.

In this paper we discuss the problem in such an indirect way that no
explicit averaging procedure is introduced, so that the results do not
depend on the averaging procedure. We simply {\it compare} the dynamical
properties of the solutions --- through the temporal behavior of their
total spatial volume $V(\tau)$, and we find a striking property of the
FLRW solution for dust matter. One of the solutions that is compared is
an FLRW solution $\dg ab^{(0)}$, and the other is an inhomogeneous
solution $\dg ab$. Synchronous and comoving conditions are assumed. The
reason we focus on the volume $V(\tau)$ is that the FLRW model has only
one dynamical variable, the scale factor $a(\tau)$, and its dynamical
content is equivalent to the volume $V_0(\tau)$. Thus, it is reasonable
to extract $V(\tau)$ also for an inhomogeneous solution and use it to
compare the solutions. To be specific, one could suppose the norm of the
difference between the solutions defined by, e.g., the $L^p$-norm
applied to the volumes,
\begin{equation}
  \label{eq:Lp}
  ||\dg ab-\dg ab^{(0)}||_p=
  \left(\int_{-\infty}^\infty |V(\tau)-V_0(\tau)|^p\d\tau\right)^{1/p},
\end{equation}
though we will not use any specific norm. The difference $\Delta
V(\tau)\equiv V(\tau)-V_0(\tau)$ itself is the object on which we focus.

Our main claim is that each element of the two-parameter set of the FLRW
solutions for dust is a {\it critical point} for $V(\tau)$, i.e., the
variations of $V(\tau)$ in the space of solutions about the FLRW
solution vanish: $\delta V(\tau)=0$. (This statement shall be made more
precise.) Due to this property the temporal behavior of an inhomogeneous
solution which is almost homogeneous and isotropic is the same as that
of an FLRW solution up to first order in the sense that $\Delta
V(\tau)\simeq \delta V(\tau)=0$. This implies that there exists a
natural correspondence between inhomogeneous solutions and the FLRW
solution. We interpret this correspondence as an {\it averaging} (up to
first order). On the other hand the second variations $\delta^2 V(\tau)$
do not vanish in general. We will also give some formulae for the second
variations.

The content of this paper is basically a direct generalization of a
previous paper, \cite{MT} where we examined the spherically symmetric
case using the exact Lema\^\i tre-Tolman-Bondi solution. The method used
to prove the claim in this paper is different from that in the previous
paper, since we cannot rely upon an exact solution. The existence of a
conserved quantity (the total number of dust particles) will be one of
the key points of our analysis. Another key is the property of a
homogeneous and isotropic (three-)metric, mainly as an Einstein metric,
i.e., that the Ricci tensor is proportional to the metric. These two
features give rise to a closed ordinary differential equation for
$\delta V(\tau)$, and which is used to show the claim. In particular, we
do not have to solve perturbation equations. The second variation
$\delta^2 V(\tau)$ is obtained in a similar way.

Our approach is not to present a complete framework for averaging, but
rather to give a transparent view of the interrelation between
inhomogeneous solutions and the FLRW solutions. This should be a helpful
guide for a detailed investigation.

In the next section we make some preparations for our analysis and
introduce some notation. In \S \ref{sec:v1} we discuss the first
variations of $V(\tau)$ mainly in the case of closed spatial
manifolds. Section \ref{sec:v2} concerns the second variations. In \S
\ref{sec:B} we comment on cases of spatial manifolds which are compact
but have boundaries. Section \ref{sec:last} is devoted to a summary.

\section{Preliminary}

Let $M$ be a 3-dimensional compact manifold which admits a constant
curvature metric $\tilde\dh ij$. We use $x^i$ to denote coordinates for
$M$. The spacetime manifold we consider is the direct product
$M\times\R$, for which we use $x^0=\tau$ to denote the coordinate for
$\R$. We consider smooth metrics on $M\times\R$ of the synchronous
form
\begin{equation}
  \label{eq:mt}
  \d s^2=\dg ab\d x^a\d x^b=\d\tau^2-\dh ij(\tau,x^i)\d x^i\d x^j.
\end{equation}
Here the spatial metric $\dh ij(\tau,x^i)$ can be considered as a
one-parameter ($\tau$) family of smooth metrics on $M$. Conversely, once
given a one-parameter family $\dh ij(\tau,x^i)$ of smooth metrics on
$M$, we can identify it with the spacetime metric \reff{eq:mt}. Thus, we
can define the space $P^*$ of smooth (synchronous) metrics on $M\times
\R$ as the set of all possible one-parameter families of smooth metrics
on $M$, $\dh ij(\tau,x^i)$. Next, let us consider Einstein's equation
for dust: $G_{ab}=8\pi\rho u_au_b$, where $G_{ab}$ is the Einstein
tensor for $\dg ab$, and $\rho$ and $u^a\equiv(\partial/\partial\tau)^a$
are, respectively, the matter density and four-velocity of dust. We
define the space of solutions $P\subset P^*$ as the set of all smooth
solutions of this equation. We distinguish $P$ and $P^*$ to clarify
whether a relation holds with dynamical equations or without them.

Since we have assumed that $M$ admits a constant curvature metric
$\tilde\dh ij$, the spacetime manifold admits a spatially homogeneous
and isotropic metric of the form
\begin{equation}
  \label{eq:defa}
  \d s^2=\dg ab^{(0)}\d x^a\d x^b=\d\tau^2-a(\tau)^2\tilde\dh ij(x^i)\d
  x^i\d x^j.
\end{equation}
The function $a(\tau)$ specifies the scale of $M$ for a slice
$\tau=$constant, and it is called the {\it scale factor}. To make the
magnitude of $a$ well-defined, we remove the scale ambiguity of
$\tilde\dh ij$ by choosing it as a standard metric, satisfying
\begin{equation}
  \label{eq:flrw0}
  {}^{(3)}\tilde R=6k,
\end{equation}
where ${}^{(3)}\tilde R$ is the scalar curvature for the standard metric
chosen, and $k=1$ for a positive curvature space, $k=-1$ for a negative
curvature space, and $k=0$ for a flat space. The constant $k$ is
referred to as the {\it curvature index}. The curvature tensor
$\sR_{ijkl}$, Ricci tensor $\sR_{ij}$, and scalar curvature $\sR$ for
the spatial metric $\dh ij=a^2\tilde\dh ij$ can be found, for example,
from the conformal invariance of the Ricci tensor $\sR_{ij}$:
\begin{equation}
  \label{eq:flrw1}
  \sR_{ij}={}^{(3)}\tilde R_{ij}=2k\tilde\dh ij=\frac{2k}{a^2}\dh ij.
\end{equation}
The curvature tensor and scalar curvature are hence
\begin{equation}
  \label{eq:flrw2}
  \sR_{ijkl}=\frac{k}{a^2}(\dh ik\dh jl-\dh il\dh jk), \; \sR=\frac{6k}{a^2}.
\end{equation}

We define the total spatial volume $V(\tau)$ for the metric \reff{eq:mt} by
\begin{equation}
  \label{eq:defV}
  V(\tau)\equiv \int_M \sqrt{h}\d^3 x,
\end{equation}
and the total particle number $E$ by
\begin{equation}
  \label{eq:defE}
  E\equiv \int_M G_{00}\sqrt{h}\d^3 x=\rcp2\int_M (\sR+K^2-\dK ij\uK
  ij)\sqrt{h}\d^3 x,
\end{equation}
where $\sR$ is the scalar curvature for the spatial metric $\dh ij$,
$\dK ij=\rcp2 \dot\dh ij$ is the extrinsic curvature, and $K\equiv \uh
ij\dK ij$, $h\equiv \det (\dh ij)$. To raise and lower the spatial
indices $i,j,\cdots$ we use $\uh ij$ and $\dh ij$, e.g., $K^i{}_j\equiv
\uh ik\dK kj$. Here dots $(\,\dot{}\,)$ represent derivatives with
respect to $\tau$.

The particle number $E$ is conserved if the dynamical and constraint
equations $G_{ij}=0=G_{0i}$ are imposed.  In fact, one can check by a
straightforward calculation that $(G_{00}\sqrt h)\dot{}=(-G_{ij}\uK
ij+\nabla^iG_{0i})\sqrt h$, implying $\dot E=0$ if we impose the
equations.

$V(\tau)$ can be regarded as a ``function-valued'' functional on $P$ (or
$P^*$), since once $\dh ij(\tau,x^i)\in P$ (or $P^*$) is specified the
function $V(\tau)$ is determined. (To express this we could add an
argument of the functional as ``$V(\tau)[\dh ij(\tau)]$,'' but for
simplicity we shall write ``$V(\tau)$.'') Similarly, $E$ is also a
functional on $P$ or $P^*$, which is real-valued on $P$, since $E$ is
conserved, but is function-valued on $P^*$.

Consider a smooth path $l$ in $P^*$, $l:\; [0,\infty) \goes P^*$. Let
$V_\epsilon(\tau)$ be the function $V(\tau)$ for $l(\epsilon)$,
where $\epsilon\in [0,\infty)$. Then, we can expand $V_\epsilon(\tau)$
at $\epsilon=0$ as
\begin{equation}
  \label{eq:Vpan}
  V_\epsilon(\tau)=V_0(\tau)+\epsilon \delta V(\tau)+\rcp2 \epsilon^2
  \delta^2 V(\tau)+\cdots,
\end{equation}
where
\begin{equation}
  \label{eq:Vvars}
  \delta V(\tau)\equiv \frac{\d V_\epsilon(\tau)}{\d
  \epsilon}\bigg|_{\epsilon=0},\; 
  \delta^2 V(\tau)\equiv \frac{\d^2 V_\epsilon(\tau)}{\d
  \epsilon^2}\bigg|_{\epsilon=0},\; \cdots.
\end{equation}
We call $\delta V(\tau)$ and $\delta^2 V(\tau)$ the {\it variation} and
{\it second variation} of $V(\tau)$, respectively. Variations of any
functional in any space are defined similarly.

Practically we will not specify any path, but we expand a variation of a
functional in terms of variations of $\dh ij$ and $\dK ij$. Thus it is
useful to introduce some notation for them.

\medskip
\noindent
{\bf Notation} Let $\delta \dh ij$ and $\delta \dK ij$ be variations of
the spatial metric and extrinsic curvature for the spacetime metric
\reff{eq:mt}. Similarly, let $\delta^2 \dh ij$ and $\delta^2 \dK ij$ be
their second variations. We use the following notation for them and
their traces throughout this paper:
\begin{equation}
  \label{eq:not1}
\gamma_{ij}\equiv\delta \dh ij,\; \gamma\equiv \uh ij \gamma_{ij},\;
\lambda_{ij}\equiv \delta \dK ij, \lambda\equiv \uh
ij\lambda_{ij},
\end{equation}
\begin{equation}
  \label{eq:not2}
  \gamsec_{ij}\equiv\delta^2\dh ij,\;
  \gamsec\equiv\uh ij\gamsec_{ij},\;
  \lamsec_{ij}\equiv\delta^2\dK ij,\; \lamsec\equiv\uh ij\lamsec_{ij}.
\end{equation}

\section{Variations of $V$}
\label{sec:v1}

Our concern is the variations of $V(\tau)$ and $E$ evaluated about the
homogeneous and isotropic metric \reff{eq:defa}.  We find a closed
relation between them if the spatial manifold $M$ is closed, but for
later convenience we present the relation in the case that $M$ is
compact first.
\begin{lemma}
  \label{th:1}
  Let $V(\tau)$ and $E$ be the functionals on $P^*$ defined by
  Eqs.\reff{eq:defV} and \reff{eq:defE}. For the variations about a
  spatially homogeneous and isotropic metric $\dg ab^{(0)}$,
  \begin{equation}
    \label{eq:fir}
    \delta E-a\delta B=\paren{\frac{k}{a^2}-
      3\paren{\frac{\dot a}{a}}^2}\delta V+
    2\frac{\dot a}{a}\delta\dot V,
  \end{equation}
  where
  \begin{equation}
    \label{eq:defdelB}
    \delta B\equiv \rcp{2a}\int_M\nabla^i(\nabla^j
    \gamma_{ij}-\nabla_i\gamma)\sqrt h\d^3 x,
  \end{equation}
  $a=a(\tau)$ is the scale factor for the homogeneous and isotropic
  spatial metric, and $k$ is the curvature index.
\end{lemma}
Here, $\delta \dot V$ is the time derivative of $\delta V$
(``$\delta$'' and `` $\dot{}$ '' commute, however), and
$\nabla_i$ represents the covariant derivative associated with (the
zeroth order of) $\dh ij$. As long as there is no confusion, we omit
superscripts (such as that on $\dh ij^{(0)}$) in reference to a
zeroth order quantity.

To prove Lemma \ref{th:1} we present some formulae, which can all be
checked by straightforward calculations.
\begin{formulae}
\label{fl:1}
The variations $\delta V$, $\delta\dot V$, and $\delta E$ (about a
generic point) are given by
\begin{equation}
  \label{eq:delV}
  \delta V=\rcp2\int_M\gamma\sqrt h\d^3 x,
\end{equation}
\begin{equation}
  \label{eq:delVd}
  \delta \dot V=\int_M \paren{\lambda+K\gamma-\dK ij\gamma^{ij}}\sqrt h\d^3 x
\end{equation}
and
\begin{eqnarray}
  \label{eq:delE}
  \delta E &=& \int_M
  \bigg[-\paren{\rcp2\sR^{ij}+K\uK ij-K^i{}_k\uK jk}\gamma_{ij} \nonumber \\
  && \quad +\rcp4\paren{\sR+K^2-\dK ij\uK ij}\gamma 
    -\paren{\uK ij-K\uh ij}\lambda_{ij}
    \nonumber \\  && \quad 
  +\rcp2 \nabla^i(\nabla^j\gamma_{ij}-\nabla_i\gamma)\bigg]\sqrt h\d^3 x.
\end{eqnarray}
\end{formulae}

{\it Proof of Lemma \ref{th:1}.} We evaluate the variations
\reff{eq:delV}, \reff{eq:delVd} and \reff{eq:delE} about a homogeneous
and isotropic metric.  The spatial part of such a metric can be written
as $\dh ij=a^2(\tau)\tilde\dh ij$, as in Eq.\reff{eq:defa}. The
extrinsic curvature becomes
\begin{equation}
  \label{eq:flrw3}
  \dK ij=\rcp2\dot\dh ij=a\dot a\tilde\dh ij=\frac{\dot a}a\dh ij.
\end{equation}
The Ricci tensor $\sR_{ij}$ is also proportional to $\dh ij$
(Eq.\reff{eq:flrw1}).  If we substitute Eqs.\reff{eq:flrw3} and
\reff{eq:flrw1} into Eqs.\reff{eq:delVd} and \reff{eq:delE}, we find
\begin{equation}
  \label{eq:delVd0}
  \delta\dot V=\int_M \paren{\lambda+\rcp 2\frac{\dot a}a\gamma}\sqrt
  h\d^3 x = \int_M\lambda\sqrt h\d^3x+\frac{\dot a}a\delta V
\end{equation}
and
\begin{equation}
  \label{eq:delE0}
  \delta E-a\delta B =\rcp2\paren{\frac{k}{a^2}-\paren{\frac{\dot
        a}a}^2}\int_M\gamma\sqrt h\d^3x+2\frac{\dot
    a}a\int_M\lambda\sqrt h\d^3x,
\end{equation}
where $\delta B$ is defined by Eq.\reff{eq:defdelB}.  The last equation
\reff{eq:delE0} implies Eq.\reff{eq:fir} if we note Eqs.\reff{eq:delV}
and \reff{eq:delVd0}. ${}_\Box$

If the spatial manifold $M$ is {\it closed} (i.e., $\partial M=\emptyset$),
the divergence term $\delta B$ vanishes, so that we have a simpler
result:
\begin{coro}
  \label{coro:1}
  Let $V(\tau)$ and $E$ be the functionals on $P^*$ defined by
  Eqs.\reff{eq:defV} and \reff{eq:defE}. If the spatial manifold $M$ is
  closed, for variations about a spatially homogeneous and isotropic
  metric $\dg ab^{(0)}$,
  \begin{equation}
    \label{eq:fir-cl}
    \delta E=\paren{\frac{k}{a^2}-3\paren{\frac{\dot a}{a}}^2}\delta
    V+2\frac{\dot a}{a}\delta\dot V,
  \end{equation}
  where $a=a(\tau)$ is the scale factor for the homogeneous and isotropic
  spatial metric, and $k$ is the curvature index.
\end{coro}

The relation \reff{eq:fir-cl} also holds for the variations taken in
$P$. In addition, we can regard $\delta E$ as a constant, since $E$ is
conserved in such a case. The scale factor $a(\tau)$ is also a definite
function which obeys the Einstein equation
\begin{equation}
  \label{eq:Eina}
  2\frac{\ddot a}a+\paren{\frac{\dot a}a}^2+\frac k{a^2}=0.
\end{equation}
Hence the relation \reff{eq:fir-cl} is a closed linear ordinary
differential equation for $\delta V$. Although we can easily find the
solution for Eq.\reff{eq:fir-cl} directly, for later convenience we
present a more general formula first:
\begin{formula}
  \label{fl:2}
  The general solution for the ordinary differential equation for $f=f(\tau)$
  \begin{equation}
    \label{eq:firgen}
    g(\tau)=\paren{\frac{k}{a^2}-3\paren{\frac{\dot a}{a}}^2}f
    +2\frac{\dot a}{a}\dot f,
  \end{equation}
  where $g(\tau)$ is a given function of $\tau$ and $a=a(\tau)$ is a
  solution for Eq.\reff{eq:Eina}, is given by
  \begin{equation}
    \label{eq:firgensol}
    f(\tau)=a^2\dot a\paren{\rcp2\int
      \frac{g(\tau)}{a{\dot a}^2}\d\tau+c},
  \end{equation}
  where $c$ is an integration constant.
\end{formula}
Applying this formula we at once obtain the following result.
\begin{lemma}
  \label{th:4}
  Let $V(\tau)$ be the functional in the space of solutions $P$ defined
  by Eq.\reff{eq:defV}. If the spatial manifold $M$ is closed,
  variations about a homogeneous and isotropic solution are given by
  \begin{equation}
    \label{eq:delVsol}
    \delta V(\tau)=a^2\dot a\paren{\frac{\delta E}2\int\frac{\d\tau}{a{\dot 
    a}^2}+\delta C},
  \end{equation}
  where $\delta E$ and $\delta C$ are constants, and $a=a(\tau)$ is a
  solution of Eq.\reff{eq:Eina}.
\end{lemma}

We present the explicit forms of the solution for completeness:
\def\tc{\tau_{0{\rm c}}}

(i) $k=1$:  $a(\tau)=a_0(1-\cos\eta),\; \tau-\tc=a_0(\eta-\sin\eta)$,
\begin{equation}
  \label{eq:delVk1}
 \delta V(\tau)= a_0^2(1-\cos\eta)\paren{\A(\eta)\frac{\delta
 E}2+\sin\eta\delta C},
\end{equation}
where $\A(\eta)\equiv (1-\cos\eta)^2-\sin\eta(\eta-\sin\eta)=
2(1-\cos\eta)-\eta\sin\eta$. 

(ii) $k=-1$: $a(\tau)=a_0(\cosh\eta-1),\; \tau-\tc=a_0(\sinh\eta-\eta)$,
\begin{equation}
  \label{eq:delVk-1}
  \delta V(\tau)= a_0^2(\cosh\eta-1)\paren{\A_-(\eta)\frac{\delta
  E}2+\sinh\eta\delta C},
\end{equation}
where $\A_-(\eta)\equiv (\cosh\eta-1)^2-\sinh\eta(\sinh\eta-\eta)=
\eta\sinh\eta-2(\cosh\eta-1)$. 

(iii) $k=0$: $a(\tau)=a_0(\tau-\tc)^{2/3}$,
\begin{equation}
  \label{eq:delVk0}
  \delta V(\tau)= \frac34\, (\tau-\tc)^2\,\delta E+
  \frac23a_0^3\,(\tau-\tc)\,\delta C.
\end{equation}

In these solutions $a_0$ and $\tc$ are constant parameters. As one can
easily see, $\tc$ is redundant as far as the FLRW solutions are
concerned, since it carries the gauge freedom associated with the choice
of the origin of time. However, we will see that $\tc$ is of some
importance in the wider context in which we are interested.

Does the constant $\delta C$ in Eq.\reff{eq:delVsol} defined as an
integration constant for the differential equation \reff{eq:fir-cl} have
an ``integral'' $C$ such that its variation coincides with the $\delta
C$ that connects variations of $E$ and $V(\tau)$ through
Eq.\reff{eq:delVsol}? We may expect that at least in a neighborhood of
the FLRW solutions there exist such a single-valued functional $C$. We
will call this functional $C$, a {\it big bang constant}. For example,
in the space of spherically symmetric solutions there exits a natural
choice of $C$ that is globally defined, \cite{MT} but in the general
case, global existence of $C$, of course, requires proof. However, we do
not discuss this problem further, since local existence is sufficient
for our purposes. We do not discuss the uniqueness of $C$ either, for
the same reason. (In fact, it would not be unique from the definition
above. A complete definition may be given after higher order variations
of $V(\tau)$ are specified.)  We will see that considering the
functional $C$ is convenient for our analysis.

\medskip

The function $V(\tau)$ for an FLRW solution, denoted $V_0(\tau)$
hereafter, depends parametrically on $a_0$ and $\tc$, and thus it varies
if $a_0$ and $\tc$ are varied (with $\tau$ fixed). For example, in the
case $k=1$ with $M\simeq S^3$, the total derivative $\d V_0$ with
respect to $a_0$ and $\tc$ is \cite{MT}
\begin{equation}
  \label{eq:dV0}
  \d V_0(\tau)=6\pi^2a_0^2(1-\cos\eta)(\A(\eta)\d a_0-\sin\eta\d\tc).
\end{equation}
Note that the time dependence of this function is the same as that of
Eq.\reff{eq:delVk1}. Similar results hold for the other cases of $k$.
To understand the implication of this fact, we first need to give some
definitions.

\medskip
\noindent
{\bf Definitions} Let $O$ be the two-dimensional space of the FLRW
solutions, and let $o(a_0,\tc)\in O$ be the FLRW solution with given
FLRW parameters $a_0$ and $\tc$. $O$ is a subspace of $P$, as well. Let
$T_{a_0,\tc}(P)$ be the tangent space at $o(a_0,\tc)\in P$. The tangent
space at $o(a_0,\tc)\in O$ can be considered as a subspace of
$T_{a_0,\tc}(P)$, and we denote it $O_{a_0,\tc}(P)$. We call
$O_{a_0,\tc}(P)$ {\it the tangent space which preserves the homogeneity
  and isotropy}.

In correspondence to $O_{a_0,\tc}(P)$ we can define a complement
$Q_{a_0,\tc}(P)$ by $T_{a_0,\tc}(P)=O_{a_0,\tc}(P)\oplus
Q_{a_0,\tc}(P)$, where $\oplus$ is the direct sum. $Q_{a_0,\tc}(P)$ is
not unique at this point, but a comparison of Eqs.\reff{eq:delVk1} and
\reff{eq:dV0} reveals that a tangent space at $o(a_0,\tc)$ spanned by
independent tangent vectors with nonvanishing $\delta E$ and $\delta C$
is naturally identified with the tangent space $O_{a_0,\tc}(P)$. This
identification makes $Q_{a_0,\tc}(P)$ unique, since we can require that
for any ${\delta\over\delta \dh ij}\in Q_{a_0,\tc}(P)$ the corresponding
variation of $V(\tau)$ vanish, $\delta V(\tau)=0$. To summarize, we
have:
\begin{theorem}
  \label{th:T1}
  Let $V(\tau)$ be the functional in the space $P$ of solutions defined
  by Eq.\reff{eq:defV}, and suppose that the spatial manifold $M$ is
  closed. Then, the tangent space $T_{a_0,\tc}(P)$ at $o(a_0,\tc)\in P$
  can be uniquely decomposed as the direct sum
  \begin{equation}
    \label{eq:dsum}
    T_{a_0,\tc}(P)=O_{a_0,\tc}(P)\oplus Q_{a_0,\tc}(P),
  \end{equation}
  where $Q_{a_0,\tc}(P)$ is the space for which the action of any
  ${\delta\over\delta \dh ij}\in Q_{a_0,\tc}(P)$ on $V(\tau)$ vanishes:
  \begin{equation}
    \label{eq:T1}
    \delta V(\tau)=0,
  \end{equation}
  and $O_{a_0,\tc}(P)$ is the tangent space which preserves the
  homogeneity and isotropy.
\end{theorem}

In the sense of Eq.\reff{eq:T1} an FLRW solution is a {\it critical
  point} for $V(\tau)$ in the space of solutions. We may interpret this
fact in connection with the averaging as follows. Let $p\in P$ be an
inhomogeneous solution which is almost homogeneous and isotropic, and
let $V_p(\tau)$ be its volume as a function of time. Then, the theorem
implies that there exist the set of values $(a_0,\tc)$ for which the
difference between the volume $V_0(\tau)$ of the FLRW solution
$o(a_0,\tc)$ and that of the inhomogeneous solution is the same up to
first order: $\Delta V(\tau)\equiv V_p(\tau)-V_0(\tau)\simeq \delta
V(\tau)=0$.

This set of values $(a_0,\tc)$ is determined by the conditions $\delta
E=0$ and $\delta C=0$. In other words, the best fit FLRW solution
$o(a_0,\tc)$ is the solution in $O$ that possesses the same particle
number $E$ and the same big bang constant $C$ as those of the solution
$p$. This correspondence $p\goes o(a_0,\tc)$ defines the map $Av: \;
P\goes O$. That is:

\noindent
{\bf Definitions} The space of solutions $P$ is foliated by the level
sets for $E$ (i.e., the sets $E=$constant), and it is also foliated by
the level sets for $C$. Let $P_{E,C}\subset P$ be the double level set
$E=$ constant ($=E$) and $C=$ constant ($=C$). In $P_{E,C}$ there exists
only one FLRW solution $o(a_0,\tc)$. The map $Av$ is defined by this
correspondence:
\begin{equation}
  \label{eq:Av}
  Av:\; P_{E,C}\goes o(a_0,\tc).
\end{equation}
We call this map the {\it averaging map} in the space of solutions.

Strictly speaking, the map $Av$ can be defined only on the domain where
the big bang constant $C$ is defined. This domain exists at least in a
neighborhood of $O$. Again, this is sufficient for our purposes.

In terms of the averaging map $Av$, we can reword Theorem \ref{th:T1} as
follows.

\begin{theorem}
  \label{th:T2}
  The variation of $V(\tau)$ along any smooth path in
  $Av^{-1}o(a_0,\tc)$ from an FLRW solution $o(a_0,\tc)$ vanishes:
  \begin{equation}
    \label{eq:firVav}
    \delta V(\tau)= 0.
  \end{equation}
\end{theorem}

\section{Second variations of $V$}
\label{sec:v2}

The variation of $V(\tau)$ about an FLRW solution is essentially zero in
the sense of Theorem \ref{th:T1} (or \ref{th:T2}). The second variation
$\delta^2 V(\tau)$ therefore gives the leading contribution to the
difference $\Delta V(\tau)$. In this section we evaluate $\delta^2
V(\tau)$ in a manner similar to the previous section.

We first obtain the following.
\begin{lemma}
\label{th:3}
Let $V(\tau)$ and $E$ be the functionals on $P^*$ defined by
Eqs.\reff{eq:defV} and \reff{eq:defE}, and suppose that the spatial
manifold $M$ is closed. For the second variations about a spatially
homogeneous and isotropic metric $\dg ab^{(0)}$,
  \begin{equation}
    \label{eq:sec}
    \delta^2 E=\paren{\frac{k}{a^2}-3\paren{\frac{\dot a}{a}}^2}\delta^2
    V+2\frac{\dot a}{a}\delta^2\dot V +\Gamma(\tau),
  \end{equation}
where $\Gamma(\tau)$ is the function of $\tau$ that is determined from
the first variation of the metric, defined by
%  \begin{eqnarray}
%    \label{eq:defGam}
%    \Gamma(\tau) &\equiv & \int_M \bigg
%    [ \daas\gamma^2-\rcp2\paren{\ka+2\daas}\gamma^{ij}\gamma_{ij}
%    \nonumber \\
%    & &+ \rcp4\gamma^{ij}(\nabla^k\nabla_k\gamma_{ij}+
%    \nabla_i\nabla_j\gamma-2\nabla_i\nabla_k\gamma_j{}^k) \nonumber \\
%    & & +\rcp4\gamma\nabla^i(\nabla^j\gamma_{ij}-\nabla_i\gamma)
%    \nonumber \\
%    & &+2\daa(\gamma^{ij}\lambda_{ij}-\gamma\lambda)
%    +(\lambda^2-\lambda^{ij}\lambda_{ij}) \bigg]\sqrt h\d^3 x.
%  \end{eqnarray}
  \begin{eqnarray}
    \label{eq:defGam}
    \Gamma(\tau) &\equiv & \int_M \bigg
    [ \daas\gamma^2-\rcp2\paren{\ka+2\daas}\gamma^{ij}\gamma_{ij}
    \nonumber \\ & &
    + \rcp4\paren{ (\nabla_i\gamma)(\nabla^i\gamma)
    -(\nabla_i\gamma_{jk})(\nabla^i\gamma^{jk}) }
    \nonumber \\ & &
    +\rcp2\paren{ (\nabla_i\gamma^i{}_k)(\nabla_j\gamma^{jk})
    - (\nabla_i\gamma)(\nabla_j\gamma^{ij}) }
    \nonumber \\ & &
    +2\daa\paren{\gamma^{ij}\lambda_{ij}-\gamma\lambda}
    +\lambda^2-\lambda^{ij}\lambda_{ij} \bigg]\sqrt h\d^3 x.
  \end{eqnarray}
  Here, $a=a(\tau)$ is the scale factor in $\dg ab^{(0)}$, and $k$ is
  the curvature index.
\end{lemma}

To prove Lemma \ref{th:3} we need to calculate the second variations
$\delta^2 E$ and $\delta^2 V(\tau)$, and the time derivative
$\delta^2\dot V$, about a generic point in $P^*$ or $P$. We obtain:
\begin{formulae}
\label{fl:3}
The second variations $\delta^2 V$, $\delta^2\dot V$ and $\delta^2 E$
are given by
\begin{equation}
  \label{eq:secV}
  \delta^2 V=
  \rcp2\int_M\paren{\gamsec-\gamma^{ij}\gamma_{ij}+\rcp2\gamma^2}
  \sqrt h\d^3x,
\end{equation}
\begin{eqnarray}
  \label{eq:secVd}
  \delta^2 \dot V &=& \int_M\bigg[ 
  \lamsec
  -2\gamma^{ij}\lambda_{ij}+\gamma\lambda
 -\uK ij\paren{\gamsec_{ij}-2\gamma_{ik}\gamma_j{}^k+\gamma\gamma_{ij}}
 \nonumber \\  & & 
  +\rcp2 K\paren{\gamsec-\gamma^{ij}\gamma_{ij}+\rcp2\gamma^2}
  \bigg] \sqrt h\d^3x,
\end{eqnarray}
and
\begin{eqnarray}
  \label{eq:secE}
  \delta^2 E &=& \int_M\bigg[
% curvature part
  \rcp2\sR_{ikjl}\gamma^{ij}\gamma^{kl}
  +\rcp2\sR_{ij}\paren{-\gamsec^{ij}+\gamma^i{}_k\gamma^{jk}-\gamma\gamma^{ij}}
  \nonumber \\ & &
  -\rcp4\sR\paren{-\gamsec+\gamma^{ij}\gamma_{ij}-\rcp2\gamma^2}
  \nonumber \\ & &
% nabla part
  +\rcp4\gamma^{ij}\paren{\nabla^k\nabla_k\gamma_{ij}+
    \nabla_i\nabla_j\gamma-2\nabla_i\nabla_k\gamma_j{}^k}
  +\rcp4\gamma\nabla^i\paren{\nabla^j\gamma_{ij}-\nabla_i\gamma}
  \nonumber \\ & &
  +\rcp2 \nabla_i(\delta\beta^i+\rcp2\gamma\beta^i)
  \nonumber \\ & &
% rest part
  +\paren{\uK ij\uK kl-\uK ik\uK jl}\gamma_{ij}\gamma_{kl}
  \nonumber \\ & &
  +\paren{K_i{}^k\dK jk-K\dK ij}
  \paren{\gamsec^{ij}-2\gamma^{il}\gamma^j{}_k+\gamma\gamma^{ij}}
  \nonumber \\ & &
  +\rcp4 \paren{K^2-\uK ij\dK ij}
  \paren{\gamsec-\gamma^{ij}\gamma_{ij}+\rcp2\gamma^2}
  \nonumber \\ & &
  -\uK ij\paren{\lamsec_{ij}-4\gamma_{ik}\lambda_j{}^k
    +2\gamma_{ij}\lambda+\gamma\lambda_{ij}}
  \nonumber \\ & &
  +K\paren{\lamsec-2\gamma^{ij}\lambda_{ij}+\gamma\lambda}
  -\lambda^{ij}\lambda_{ij}
  +\lambda^2
  \bigg]\sqrt h\d^3x,
\end{eqnarray}
where
\begin{equation}
  \label{eq:defbeta}
  \beta_i\equiv\nabla^j\gamma_{ij}-\nabla_i\gamma
\end{equation}
and $\delta\beta^i$ is the variation of $\beta^i=\uh ij\beta_j$.
\end{formulae}

{\it Proof of Lemma \ref{th:3}.} We evaluate Eqs.\reff{eq:secVd} and
\reff{eq:secE} about the FLRW solution, using Eqs.\reff{eq:flrw1},
\reff{eq:flrw2}, and \reff{eq:flrw3}. The result for $\delta^2\dot V$ is
\begin{eqnarray}
  \label{eq:secVd0}
  \delta^2\dot V &=& \int_M\paren{ \lamsec
  -2\gamma^{ij}\lambda_{ij}+\gamma\lambda}\sqrt h\d^3 x
% \nonumber \\  & & 
  +\frac{\dot a}{2a}\int_M\paren{\gamsec+\gamma^{ij}\gamma_{ij}-\rcp2 
  \gamma^2}\sqrt h\d^3 x \nonumber \\
  &=& 
 \int_M\paren{ \lamsec-2\gamma^{ij}\lambda_{ij}+\gamma\lambda} \sqrt
  h\d^3 x
  +\frac{\dot a}{a}\delta^2 V 
 \nonumber \\  & & \hspace{10em}
+\frac{\dot a}{a}\int_M
  \paren{\gamma^{ij}\gamma_{ij}-\rcp2 \gamma^2}\sqrt h\d^3 x ,
\end{eqnarray}
where in the last equality we have used Eq.\reff{eq:secV}.
A straightforward evaluation of $\delta^2 E$ gives
\begin{eqnarray}
  \label{eq:secE0}
  \delta^2 E &=& \rcp2\paren{\ka-\daas}\int_M\gamsec\sqrt h\d^3x
  +2\daa\int_M\lamsec\sqrt h\d^3x
  \nonumber \\ & &
  +\int_M\bigg[ \rcp4\paren{\ka-\daas}\gamma^2
    -\paren{\ka+\frac32\daas}\gamma^{ij}\gamma_{ij}
  \nonumber \\ & &
  +\rcp4\gamma^{ij}(\nabla^k\nabla_k\gamma_{ij}+
    \nabla_i\nabla_j\gamma-2\nabla_i\nabla_k\gamma_j{}^k)
%  \nonumber \\ & & 
  +\rcp4\gamma\nabla^i(\nabla^j\gamma_{ij}-\nabla_i\gamma)
  \nonumber \\ & &
  -2\daa\gamma^{ij}\lambda_{ij}
    +(\lambda^2-\lambda^{ij}\lambda_{ij}) \bigg]\sqrt h\d^3 x.
\end{eqnarray}
We have dropped the divergence term
$\rcp2\int\nabla_i(\delta\beta^i+\rcp2\gamma\beta^i)\sqrt h\d^3 x$.
Finally, applying integration by parts to the terms containing covariant 
derivatives, and using Eqs.\reff{eq:secV} and \reff{eq:secVd0}, we obtain
Eq.\reff{eq:sec} with Eq.\reff{eq:defGam}.
${}_\Box$

If the variations are taken in $P$, $\delta^2 E$ is a constant. The
function $\Gamma(\tau)$ is determined from the first variation of the
metric, and thus it can be thought of as a given function.  Applying
Formula \ref{fl:2}, we obtain
\begin{lemma}
  Let $V(\tau)$ be the functional in the space of solutions $P$ defined
  by Eq.\reff{eq:defV}, and suppose that the spatial manifold $M$ is
  closed. Then, the second variation about a homogeneous and
  isotropic solution is given by
  \begin{equation}
    \label{eq:secVsol}
    \delta^2 V(\tau)=a^2\dot a\paren{\frac{\delta^2
        E}2\int\frac{\d\tau}{a{\dot a}^2}+\delta^2 C} -\frac{a^2\dot
      a}{2}\int\frac{\Gamma(\tau)}{a\dot a^2}\d\tau,
  \end{equation}
  where $\delta^2 E$ and $\delta^2 C$ are constants, $a=a(\tau)$ is a
  solution of Eq.\reff{eq:Eina}, and $\Gamma(\tau)$ is defined by
  Eq.\reff{eq:defGam}. 
\end{lemma}

The integration constant $\delta^2 C$ appearing above is naturally
identified with the variation of $\delta C$ appearing in Lemma
\ref{th:4}, since their coefficients are the same. The constant
$\delta^2 C$ is therefore the second variation of the big bang constant
$C$. Our conclusion in this section is the following (cf. Theorem
\ref{th:T2}):

\begin{theorem}
  \label{th:T3}
  The second variation of $V(\tau)$ along a smooth path in
  $Av^{-1}o(a_0,\tc)$ from an FLRW solution $o(a_0,\tc)$ is given by
  \begin{equation}
    \label{eq:secVav}
    \delta^2 V(\tau)= -\frac{a^2\dot a}{2}\int\frac{\Gamma(\tau)}{a\dot
    a^2}\d\tau,
  \end{equation}
  where $a=a(\tau)$ is a solution of Eq.\reff{eq:Eina}, and
  $\Gamma(\tau)$ is defined by Eq.\reff{eq:defGam}.
\end{theorem}

\section{Comment on the boundary effect}
\label{sec:B}

In the previous two sections we have mainly considered the cases of
spatial manifolds being closed, i.e. $\partial M=\emptyset$. Such a case
is clearest from a theoretical point of view, since we do not need
information on boundaries. However, we may also be interested in a
compact spatial manifold with boundaries, for example, a spatial
manifold inside the horizon. If we restrict to the case in which the
boundaries are comoving so that the particle number $E$ is conserved, we
can apply a procedure similar to the ones used in the previous sections.
See Lemma \ref{th:1}. Now $\delta B$ does not vanish, but it can be
shown that it is {\it constant}. In fact, the linearized momentum
constraint equation proves $\delta\dot B=0$. Thus we can again integrate
Eq.\reff{eq:fir} using Formula \ref{fl:2}, and we obtain
\begin{equation}
  \label{eq:delVsolB}
    \delta V(\tau)=a^2\dot a\paren{\frac{\delta E}2\int\frac{\d\tau}{a{\dot 
    a}^2}-\frac{\delta B}{2}\int\frac{\d\tau}{\dot a^2}+\delta C},
\end{equation}
instead of Eq.\reff{eq:delVsol}, where $\delta E$, $\delta B$, and
$\delta C$ are constants.

The explicit forms of the term for the boundary effect
\begin{equation}
  \label{eq:VB}
  \delta V_B(\tau)\equiv -\frac{\delta B}{2}a^2\dot
  a\int\frac{\d\tau}{\dot a^2}
\end{equation}
are the following.

(i) For $k=1$,
\begin{equation}
  \label{eq:delVBk1}
 \delta V_B(\tau)= -\frac{\delta
 B}{2}a_0^3(1-\cos\eta)\paren{4(1-\cos\eta)+\sin\eta(\sin\eta-3\eta)}.
\end{equation}

(ii) For $k=-1$,
\begin{equation}
  \label{eq:delVBk-1}
 \delta V_B(\tau)= -\frac{\delta
 B}{2}a_0^3(\cosh\eta-1)\paren{4(\cosh\eta-1)-\sinh\eta(3\eta-\sinh\eta)}.
\end{equation}

(iii) For $k=0$,
\begin{equation}
  \label{eq:delVBk0}
  \delta V_B(\tau)= -\frac{\delta B}{2}\frac{9}{10}\, a_0(\tau-\tc)^\frac83.
\end{equation}

The mode corresponding to $\delta V_B(\tau)$ is not obtained from the
space of FLRW solutions.

\section{Summary}
\label{sec:last}

We have succeeded in obtaining the variations $\delta V(\tau)$ of the
spatial volume about an FLRW solution in the space of solutions for a
dust system (Lemma \ref{th:4}). It is noteworthy that this has been done
without solving linearized Einstein equations, and the result contains
all the cases of curvature index $k$. Also, we remark that we did {\it
  not} split apart $\gamma_{ij}=\delta\dh ij$ into tensor, vector, and
scalar parts, as is done in usual cosmological perturbation theory (e.g.
Ref.~\citen{cosp}). Hence our result does not depend on such a part.

It is, however, more surprising that only two modes (patterns of
time-dependence) appear in $\delta V(\tau)$, in spite of the fact that
the space of solutions is infinite dimensional. This implies that almost
all possible modes vanish. The origin of the two modes is that the FLRW
solution is a two-parameter solution. In fact, explicit calculations
(see Eq.\reff{eq:dV0}) reveal that the two modes caused by variations
with respect to the FLRW parameters $(a_0,\tc)$ coincide with those
found in the (full) space of solutions. Clearly, these modes are not of
essential significance in themselves. Instead, their existence tells us
that we should divide the space of solutions in such a way that in each
piece there exists only one FLRW solution. As we have seen, this can be
done by considering the foliation defined by the level sets $E=$ const
and $C=$ const. If the variations are taken in such a set (leaf) (i.e.,
if $\delta E=0$ and $\delta C=0$), the variation of volume $\delta
V(\tau)$ completely vanishes (Theorem \ref{th:T1} and \ref{th:T2}). Each
FLRW solution is therefore a critical point in the leaf that contains
only one FLRW solution. Every inhomogeneous solution in such a leaf is
therefore naturally related to the unique FLRW solution (cf.
Eq.\reff{eq:Av}), and this correspondence has defined our averaging (up
to first order).

The second variations do not vanish in general (See
Eq.\reff{eq:secVav}). We cannot therefore define an averaging at second
order in the space of solutions. One may need to generalize the space of
solutions to define it. This remains as a future work.

\section*{Acknowledgments}

The author wishes to thank Professor H.~Kodama for helpful discussions.
He also acknowledges financial support from the Japan Society for the
Promotion of Science and the Ministry of Education, Science and Culture.

%\begin{references}

\end{document}